\documentclass[aps,pra,lengthcheck,superscriptaddress,reprint,showpacs,longbibliography,nofootinbib,amsmath,amssymb]{revtex4-1}
\usepackage[utf8]{inputenc}
\usepackage[T1]{fontenc}

\usepackage{amssymb}
\usepackage{graphicx}
\usepackage{grffile}
\usepackage[percent]{overpic}
\usepackage{microtype}
\usepackage[usenames,dvipsnames]{xcolor}
\usepackage[colorlinks,linkcolor=blue,citecolor=blue,urlcolor=blue]{hyperref}

\def\equationautorefname~#1\null{Eq.~(#1)\null}

\DeclareMathOperator{\Tr}{Tr}

\renewcommand{\Re}{\operatorname{Re}}
\renewcommand{\Im}{\operatorname{Im}}

\begin{document}

\title{Optomechanical heat transfer between molecules in a nanoplasmonic cavity}

\author{S. Mahmoud Ashrafi}
\affiliation{Department of Physics, Tarbiat Modares University, Tehran, Iran}
\author{R. Malekfar}
\affiliation{Department of Physics, Tarbiat Modares University, Tehran, Iran}
\author{A. R. Bahrampour}
\affiliation{Department of Physics, Sharif University of Technology, Tehran, Iran}
\author{Johannes Feist}
\email{johannes.feist@uam.es}
\affiliation{Departamento de Física Teórica de la Materia Condensada and
Condensed Matter Physics Center (IFIMAC), Universidad Autónoma de Madrid,
E-28049 Madrid, Spain}

\begin{abstract}
  We explore whether localized surface plasmon polariton modes can transfer heat
  between molecules placed in the hot spot of a nanoplasmonic cavity through
  optomechanical interaction with the molecular vibrations. We demonstrate that
  external driving of the plasmon resonance indeed induces an effective
  molecule-molecule interaction corresponding to a new heat transfer mechanism,
  which can even be more effective in cooling the hotter molecule than its
  heating due to the vibrational pumping by the plasmon. This novel mechanism
  allows to actively control the rate of heat flow between molecules through the
  intensity and frequency of the driving laser.
\end{abstract}

\maketitle

\section{Introduction}
Achieving thermal control of molecular systems is a topic of current interest in
various fields, such as quantum thermodynamics, quantum biology and quantum
chemistry~\cite{Leitner2009, OReilly2014, Chen2017Molecular, Katz2016, Lee2007,
Wu2011, Nalbach2010, Shapiro2012}. In order to achieve this, it is necessary to
gain a fundamental understanding of the transfer of heat, and energy in general,
between molecules. In addition to well-known mechanisms like advection,
convection, conduction and radiation, which are responsible for the majority of
heat transfer on macroscopic scales, additional mechanisms can play an important
role in microscopic and/or quantum systems. Some examples of such mechanisms are
seen in single-atom junctions~\cite{Cui2017Quantized}, the driven
non-equilibrium spin-boson model~\cite{Wang2017Unifying}, or in spatially
separated molecules entangled through strong coupling to cavity
modes~\cite{Feist2015,Zhong2017}. In particular, optomechanical systems in which
photonic modes are coupled to mechanical degrees of freedom, have been studied
in detail in this context~\cite{Farman2014,Degenfeld-Schonburg2016,Xu2017}. 

In this article, we demonstrate that localized surface plasmon polaritons modes
can transfer heat between molecules placed in the hot spot of a plasmonic
cavity. This transfer is mediated through the molecular optomechanical
interaction between electromagnetic modes and molecular vibrations, and is made
possible by the fact that such systems can reach a high optomechanical coupling
rate within the resolved-sideband limit~\cite{Roelli2016, Benz2016,
Schmidt2016Quantum, Schmidt2016Nanocavities, Schmidt2017, Lombardi2018}. While
``traditional'' optomechanics is concerned with the interaction of
electromagnetic modes with macroscopic mechanical resonators (often the mirrors
forming the cavity), with implementations in diverse setups such as optical
Fabry-Perot cavities, optomechanical crystals, microwave LC-circuits or
membrane-in-the-middle setups~\cite{Aspelmeyer2014}, it was recently shown that
the interaction between vibrational modes of single molecules and plasmonic
cavities can be understood within the same framework, as first applied in the
context of Surface-Enhanced Raman Scattering
(SERS)~\cite{Roelli2016,Schmidt2016Nanocavities}. The optomechanical interaction
then occurs between two non-resonant (approximately) harmonic oscillators, a
localized surface plasmon resonance (LSPR) functioning as the optical mode, and
nuclear motion in the molecule functioning as the mechanical resonator, with
vibrational displacement of the nuclei causing a dispersive shift of the LSPR
frequency. While both quantum and classical molecular optomechanics (QMO and
CMO, respectively) correctly describe elementary characteristics of Raman
scattering, namely the dependence of Raman signal on the power and frequency of
the incident laser, and on the temperature~\cite{Schmidt2017}, QMO predicts
several phenomena that are not seen within CMO, such as dynamical back-action
amplification of the vibrational mode~\cite{Schmidt2016Nanocavities}, and gives
access to non-classical observables such as correlations of the emitted
photons~\cite{Schmidt2016Quantum}. Furthermore, it allows to distinguish two
adjacent molecules with similar chemical structure by the splitting of the
transparency peak~\cite{Liu2017Coupled}.

\begin{figure}
  \resizebox{\linewidth}{!}{
    \begin{overpic}[height=4cm]{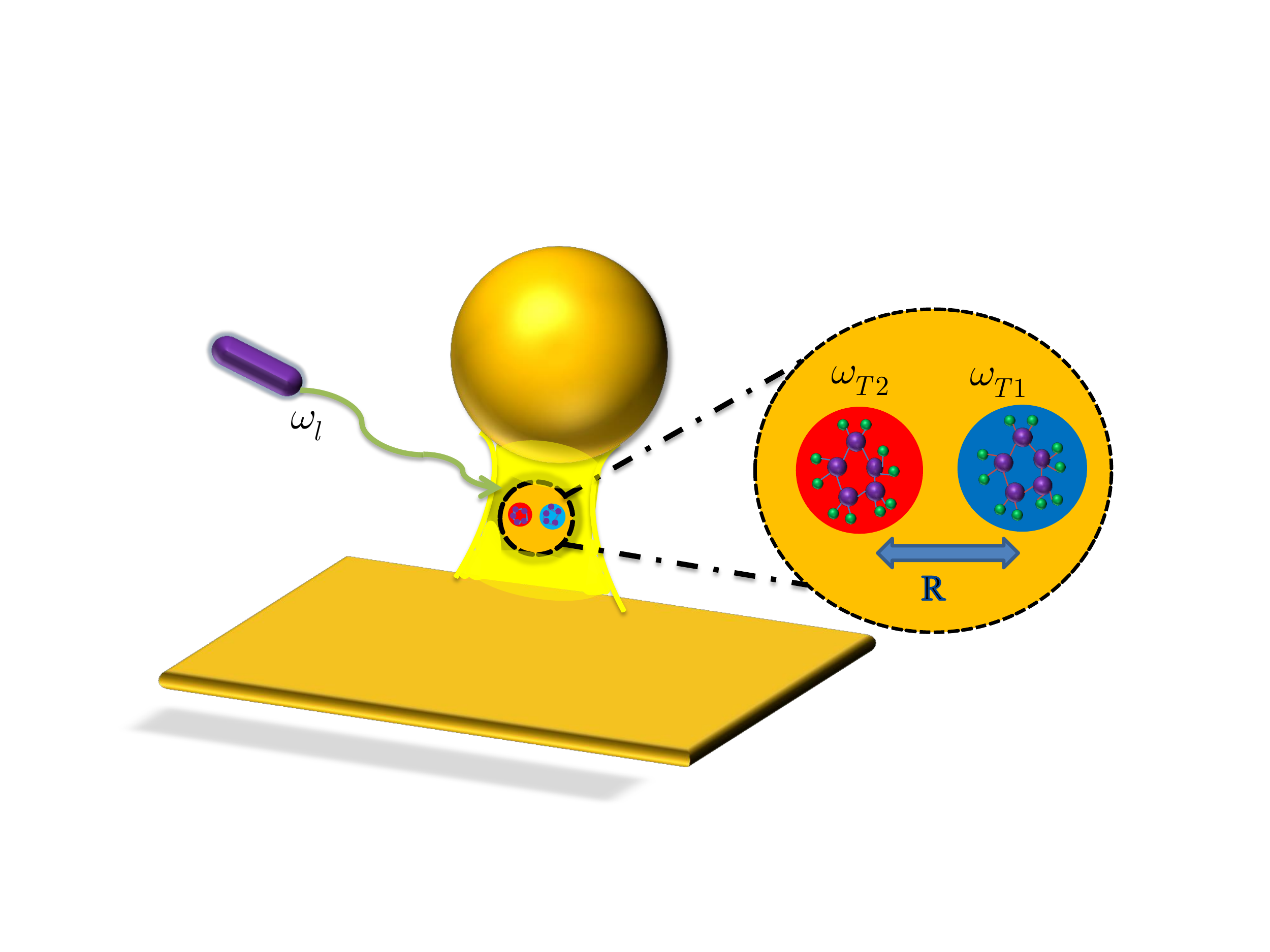} \put(0,55){(a)} \end{overpic}
    \hspace{0.5cm}
    \begin{overpic}[height=4cm]{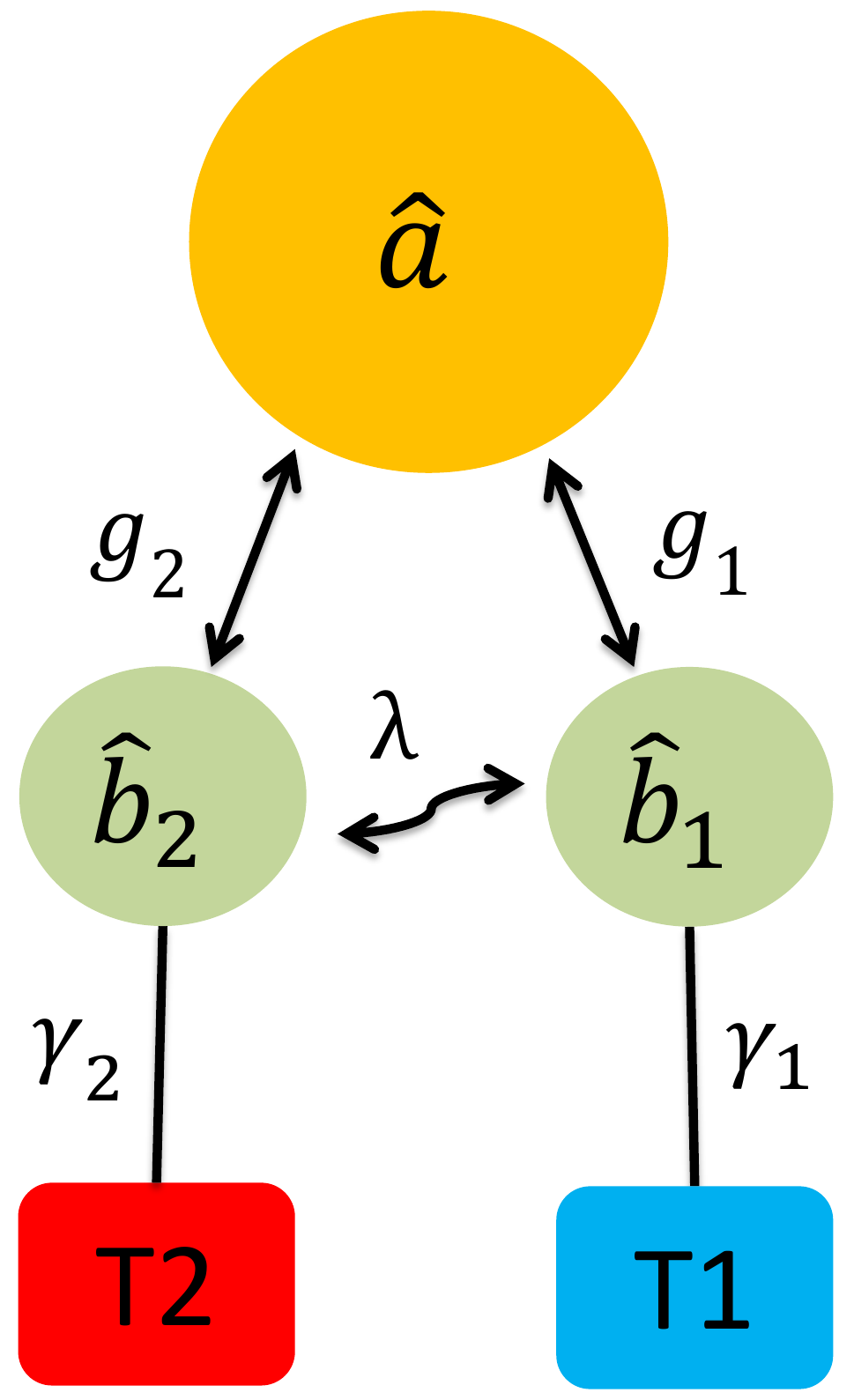} \put(0,95){(b)} \end{overpic}
  } \caption{(a) Sketch of the system of a ``cold'' (blue, $T_1$) and ``hot''
  (red, $T_2$) molecule coupled to a plasmonic resonance. (b) Schematic of the
  model and the relevant parameters (see text for details).}
  \label{fig:1}
\end{figure}

We here first demonstrate heat transfer between the vibrational degrees of
freedom of two molecules interacting with a single plasmonic cavity mode that is
driven by an external laser, as sketched in \autoref{fig:1}. We then show that
for realistic parameters, this heat transfer can become efficient enough to
offset the single-molecule plasmon-induced heating and lead to effective cooling
of the hotter molecule. Furthermore, we show that for the case of slightly
different vibrational frequencies, a competition between coherent and incoherent
coupling terms leads to an asymmetric lineshape, with the surprising result that
the colder molecule becomes further cooled down as the vibrational frequencies
of the molecules approach each other, which would normally lead to more
efficient heat transfer.

The paper is organized as follows: in \autoref{sec:theory}, we introduce the QMO
model for the system and describe our theoretical approach. Through adiabatic
elimination (\autoref{sec:adiabatic_elimination}) of the cavity mode, we derive
a simplified model in which the two molecular vibrations are directly coupled
both through coherent and incoherent coupling terms. In \autoref{sec:results},
we present our results for heat transfer between the molecules, both for the
symmetric case of identical molecules (\autoref{sec:sym}), and the asymmetric
case where the molecules and thus their vibrational modes are different
(\autoref{sec:asym}). In \autoref{sec:psd}, we analyze the results in more
detail based on the power spectral density of the oscillations. We conclude with
a summary and discussion of the results in \autoref{sec:summary}. In the
following, we use atomic units ($\hbar=1$) unless otherwise stated.

\section{Theoretical framework and model}\label{sec:theory}
The aim of our work is to investigate the heat transfer between the vibrational
modes of two molecules placed in the hot spot of a LSPR mode that mediates the
heat transfer, as shown in \autoref{fig:1}. The theoretical approach then
follows from a straightforward extension of single-molecule
descriptions~\cite{Roelli2016,Schmidt2017}, sketched in the following. We assume
that the plasmonic resonance is far-detuned from any transition within the
molecule, and treat a single vibrational mode (approximated by a harmonic
oscillator) in each molecule. The interaction between a molecular vibration and
the quantized LSPR mode can then be described by the interaction Hamiltonian
\begin{equation}\label{eq:H_ind}
H_{\mathrm{int}} = -\frac{1}{2}\hat{P}(t) \cdot \hat{E}(t),
\end{equation}
in which $\hat{P}(t)$ and $\hat{E}(t)$ are the molecular polarization and LSPR
electric field operator, respectively. Under the assumption that the plasmonic
resonance frequency is much larger than the molecular vibrational frequency,
$\omega_{c} \gg \omega_{m}$, this can be expressed through a ``standard''
optomechanical interaction~\cite{Roelli2016,Schmidt2017}, given by
\begin{equation}
  H_{\text{int}} = - g \hat{a}^{\dagger}\hat{a}(\hat{b} + \hat{b}^{\dagger}).
\end{equation}
Here, $\hat{a}$ and $\hat{b}$ are the annihilation operators for optical and
vibrational modes, respectively, and $g =
\frac{Q_{k}^{0}R_{k}\omega_{c}}{2\varepsilon_{0}\varepsilon V}$ is the
optomechanical coupling constant. This coupling depends on both the properties
of the nanocavity (permittivity of the surrounding medium
$\varepsilon_{0}\varepsilon$, effective mode volume $V$ and central frequency
$\omega_{c}$) and of the molecule (isotropic Raman tensor $R_{k}$ and zero-point
amplitude of the vibration $Q_{k}^{0}$). We now assume that we have two
molecules, separated by a distance $R$, as well as an external laser driving the
LSPR mode. The full Hamiltonian in the rotating frame of the laser is then given
by
\begin{multline}\label{eq:hamiltonian}
  \hat{H} = \delta_{0} \hat{a}^{\dagger} \hat{a} + \omega_{1} \hat{b}_{1}^{\dagger}\hat{b}_{1} + \omega_{2}\hat{b}_{2}^{\dagger}\hat{b}_{2} \\
  - g_{1}\hat{a}^{\dagger}\hat{a}\left( \hat{b}_{1}^{\dagger} + \hat{b}_{1} \right) - g_{2}\hat{a}^{\dagger}\hat{a}\left( \hat{b}_{2}^{\dagger} + \hat{b}_{2} \right) \\
  - \lambda\left( \hat{b}_{1}^{\dagger} + \hat{b}_{1} \right)\left( \hat{b}_{2}^{\dagger} + \hat{b}_{2} \right) + i\Omega\left( \hat{a}^{\dagger} - \hat{a} \right),
\end{multline}
where $\delta_{0} = \omega_{c} - \omega_{l}$ is plasmon-pump detuning and
$\Omega$ determines the laser pump intensity, while
\begin{equation}
  \lambda = \frac{\vec{d}_1 \cdot \vec{d}_2 - 3(\vec{d}_1\cdot\hat{n}) (\vec{d}_2\cdot\hat{n})}{4\pi\epsilon_0 R^3}
\end{equation}
is the dipole-dipole coupling constant between the molecules, with $\vec{d}_i$
the vibrational transition dipole moment of molecule $i$, and $R \hat{n} =
\vec{r}_2 - \vec{r}_1$ the vector connecting the two molecules. We note that to
lowest order, the vibrational transition dipole moment $\vec d_i$ depends on the
derivative of the permanent molecular dipole moment as a function of the
relevant nuclear coordinate $Q$, i.e., $\vec\mu(Q)\approx \vec\mu(0) +
\vec\mu'(0) Q = \vec\mu(0) + \vec{d} (b+b^\dagger)$, while the optomechanical
interaction depends on the derivative of the molecular polarizability tensor
$\alpha(Q) \approx \alpha(0) + R_k Q$ (assumed isotropic here for simplicity),
and the two are thus only indirectly related. We also note that both
$\vec\mu(0)$ and $\alpha(0)$ can be removed from the Hamiltonian by suitable
shifts of the equilibrium positions and frequencies. Furthermore, we assume that
the molecules are coupled to independent heat baths at temperatures $T_1$ and
$T_2$, as sketched in \autoref{fig:1}(b). Plasmonic hotspots are typically very
small ($\ll 100$~nm), such that having different local temperatures for the
molecules would require a very local source of heating. This could be achieved,
e.g., with nanometric tips used as near-field thermal probes and for radiative
heat transfer
experiments~\cite{Desiatov2014,Kim2015b,Cui2017Study,Kloppstech2017}, or through
frequency-selective resonant laser heating of different bath molecules (e.g., by
using DNA origami to precisely control molecular
positions~\cite{Chikkaraddy2018}). Alternatively, it would be possible to place
the two molecules in different hot spots of the same LSPR mode, as, e.g.,
provided by triangular plasmonic nanoparticles~\cite{Zengin2015}.

In addition to the coherent dynamics described by \autoref{eq:hamiltonian}, we
include the coupling to external baths within a Lindblad master equation
description~\cite{Gardiner2004,Breuer2007}:
\begin{equation}\label{eq:master}
  \dot\rho = \frac{1}{i} [\hat{H},\rho] + L_{\hat{a}}[\rho] + L_{\hat{b}_{1}}[\rho] + L_{\hat{b}_{2}}[\rho],
\end{equation}
where
\begin{subequations}
\begin{align}
  L_{\hat{b}_{i}}[\rho] &= \gamma_{i} (\bar{n}_{i} + 1) D_{\hat{b}_{i}}[\rho] 
                         + \gamma_{i}  \bar{n}_{i}      D_{\hat{b}_{i}^{\dagger}}[\rho],\\
  L_{\hat{a}}[\rho] &= \kappa D_{\hat{a}}[\rho],
\end{align}
\end{subequations}
where $\kappa$ describes the decay of the plasmon, and $\gamma_{1}$,
$\gamma_{2}$ are the coupling rates of the first and second molecule to their
respective thermal baths, while $D_{\hat{C}}[\rho]$ is a standard Lindblad
superoperator,
\begin{equation}
  D_{\hat{C}}[\rho] = \hat{C}\rho\hat{C}^{\dagger} - \frac12 \left(\hat{C}^{\dagger}\hat{C}\rho + \rho\hat{C}^{\dagger}\hat{C}\right).
\end{equation}
The temperature of the two baths is encoded in the mean phonon occupation
numbers ($\bar{n}_{1}$, $\bar{n}_{2}$), given by~\cite{Gardiner2004,Breuer2007}
\begin{equation}\label{eq:mean_occ}
\bar{n}_{i} = \frac{1}{\exp\big(\frac{\omega_{i}}{k_{B}T_{i}}\big) - 1},
\end{equation}
where $k_{B}$ is the Boltzmann constant. 
If the molecule-molecule and molecule-plasmon interaction is negligible, each
molecule will reach thermal equilibrium with its bath, with the populations of
the vibrational levels decaying exponentially following a Boltzmann
distribution. The average phonon number, i.e., the expectation value $n_i =
\langle \hat{b}_{i}^{\dagger} \hat{b}_{i} \rangle_\text{ss} =
\Tr(\hat{b}_{i}^{\dagger}\hat{b}_{i} \rho_{\text{ss}})$, where
$\rho_{\text{ss}}$ is the steady-state density matrix, then becomes equal to
$\bar n_i$.

The dipole-dipole and optomechanical interaction between molecules can modify
the temperature, and more generally, the steady-state distributions. We then
define an effective temperature based on the average phonon number, i.e.,
\begin{equation}
T_{i}^{\text{eff}} = \frac{\omega_{i}}{k_{B}\log\left( 1 + 1/n_{i} \right)}.
\end{equation}
In order for this effective temperature to correspond to a physical temperature,
the population of the separate levels should again follow a thermal
distribution. We have checked for all the results presented below that this is
indeed the case, i.e., that the steady-state distributions of the phonon
populations are well-approximated by thermal Boltzmann distributions, and the
effective temperatures obtained can thus indeed be interpreted as the
steady-state physical temperatures of the respective vibrational modes.

Finally, we mention that all the numerical results shown below are obtained
using the open-source QuTiP package~\cite{Johansson2012,Johansson2013}. In the
numerical calculations, we have used a cutoff of $N=6$ for the maximum phonon
and photon numbers. We have checked that this provides converged results for all
the parameters considered below.

\subsection{Adiabatic elimination of the cavity mode}\label{sec:adiabatic_elimination}
In order to analyze the numerical results below and gain more physical insight,
we perform adiabatic elimination of the plasmon mode, which leads to an
effective Hamiltonian for the two vibrational modes. Our derivation generalizes
results obtained for the heat transfer between identical harmonic
oscillators~\cite{Xuereb2015} by allowing different frequencies and coupling
strengths for the two oscillators. We neglect direct dipole-dipole interactions
in the following derivation.

To perform the adiabatic elimination, we work in the linearized limit of
optomechanics~\cite{Aspelmeyer2014}. This amounts to displacing the oscillators,
$\hat{a} \to \alpha + \hat{a}$, $\hat{b}_i \to \beta_i + \hat{b}_i$, where
$\alpha \approx \frac{\Omega}{\kappa/2 + i \Delta}$ and $\beta_i \approx
\frac{g_i |\alpha|^2}{\omega_i - i\gamma_i/2}$ are the steady-state expectation
values of $\hat{a}$ and $\hat{b}$, respectively, with $\Delta = \delta_{0} -
2g_1\Re\beta_1 - 2g_2\Re\beta_2$. After the transformation, $\langle
\hat{a}\rangle = \langle \hat{b}_1\rangle = \langle \hat{b}_2\rangle =0$, and
the driving term $\propto\Omega$ disappears. By dropping quadratic operator
terms in addition, this gives
\begin{multline}
\!\!\!\hat{H}\approx\Delta \hat{a}^\dagger \hat{a} 
  + \sum_i \left[ \omega_i \hat{b}_i^\dagger \hat{b}_i 
  - (\alpha \hat{a}^\dagger + \alpha^* \hat{a}) g_i (\hat{b}_i+\hat{b}_i^\dagger) \right].
\end{multline}
Inserting this in~\autoref{eq:master} and following the approach
of~\cite{Xuereb2015} to adiabatically eliminate the plasmon mode, we get after
some algebra
\begin{multline}\label{eq:master_adiab}
\dot\rho_{\mathrm{s}} = -i[\hat{H}_{\mathrm{s}}, \rho_{\mathrm{s}}] + L_{\hat{b}_{1}}[\rho_{\mathrm{s}}] + L_{\hat{b}_{2}}[\rho_{\mathrm{s}}]\\
 + |\alpha|^2 \left(\hat{B} \rho_{\mathrm{s}} \hat{B}_\omega - \rho_{\mathrm{s}} \hat{B}_\omega \hat{B} + \hat{B}_\omega^\dagger \rho_{\mathrm{s}} \hat{B} - \hat{B} \hat{B}_\omega^\dagger \rho_{\mathrm{s}} \right),
\end{multline}
where $\hat{H}_{\mathrm{s}} = \sum_i \omega_i \hat{b}_i^\dagger \hat{b}_i$, $\rho_{\mathrm{s}} =
\Tr_a\rho$ and 
\begin{align}
  \hat{B} &= g_1 (\hat{b}_1 + \hat{b}_1^\dagger) + g_2 (\hat{b}_2 + \hat{b}_2^\dagger)\\
  \hat{B}_\omega &= g_1 S_1 \hat{b}_1 + g_1 S_{-1} \hat{b}_1^\dagger + g_2 S_2 \hat{b}_2 + g_2 S_{-2} \hat{b}_2^\dagger.
\end{align}
Here, we have defined $S_i = S(\omega_i)$ and $S_{-i} = S(-\omega_i)$, with
$S(\omega) = [\kappa/2 - i(\Delta+\omega)]^{-1}$. We note that
\autoref{eq:master_adiab} is not in canonical Lindblad form~\cite{Lindblad1976},
but could be rewritten in this form and expressed using the original system
operators $\hat{b}_i$, $\hat{b}_i^\dagger$. The resulting expression is unwieldy
and is not shown here. It can be simplified by performing the rotating wave
approximation (RWA), i.e., only keeping slowly rotating terms that conserve the
number of vibrations. This gives
\begin{gather}\label{eq:master_adiab_RWA}
  \dot\rho_{\mathrm{s}} = -i[H_{\mathrm{s}} + H_{\mathrm{ad}}^{\mathrm{RWA}}, \rho_{\mathrm{s}}] + \sum_i L_{\hat{b}_{i}}[\rho_{\mathrm{s}}] + L_{\mathrm{ad}}^{\mathrm{RWA}}[\rho_{\mathrm{s}}],
\end{gather}
where the plasmon-induced coherent interaction is
\begin{subequations}
\begin{align}
  H_{\mathrm{ad}}^{\mathrm{RWA}} &= \sum_{i} \delta\omega_i \hat{b}_{i}^{\dagger} \hat{b}_{i} + (\Lambda \hat{b}_{1}^{\dagger} \hat{b}_{2} + \mathrm{H.c.}),\\
  \delta\omega_i &= -|\alpha|^2 g_{i}^{2} \Im(S_{i}+S_{-i}),\\
  \Lambda &= \frac{i}{2} |\alpha|^2 g_{1} g_{2} \left(S_{2} + S_{-1} - S_{1}^* - S_{-2}^*\right),
  \label{eq:Lambda}
\end{align}
\end{subequations}
which in addition to energy shifts of the oscillators gives an effective
coupling $\Lambda$. We note that for the symmetric case of identical molecules,
$\omega_2=\omega_1$, $g_2=g_1$, the new Hamiltonian can be written as
\begin{equation}
  H_{\mathrm{ad}}^{\mathrm{RWA}} = -2|\alpha|^2 g_1^2 \Im(S_1+S_{-1}) \hat{b}_c^\dagger \hat{b}_c,
\end{equation}
where $\hat{b}_c = (\hat{b}_1+\hat{b}_2)/\sqrt{2}$ is the center-of-mass mode of
the molecular vibrations.

The incoherent contribution to the dynamics is
\begin{equation}\label{eq:Lad_RWA}
  L_{\mathrm{ad}}^{\mathrm{RWA}}[\rho_{\mathrm{s}}] = \sum_{i,j} A_{ij}^{-} F_{\hat{b}_i,\hat{b}_j}[\rho_{\mathrm{s}}] 
    + A_{ij}^{+} F_{\hat{b}_i^\dagger,\hat{b}_j^\dagger}[\rho_{\mathrm{s}}],
\end{equation}
with $A_{ij}^\pm  = |\alpha|^2 g_i g_j (S_{\pm i}^* + S_{\pm j})$ and
$F_{c,d}[\rho] = c\rho d^\dagger - \frac12 \{d^\dagger c, \rho\}$.
\autoref{eq:Lad_RWA} can be brought into standard Lindblad form by diagonalizing
the matrices $A_{ij}^\pm$. We here only show the result for the symmetric case
($\omega_2=\omega_1$, $g_2=g_1$), for which a simple analytical result is obtained:
\begin{equation}\label{eq:Lad_RWAsym}
  L_{\mathrm{ad}}^{\mathrm{sym}}[\rho_{\mathrm{s}}] = 4|\alpha|^2 g_1^2 \left( \Re\!S_{-1} D_{\hat{b}_c}[\rho_{\mathrm{s}}] + \Re\!S_{1} D_{\hat{b}_c^\dagger}[\rho_{\mathrm{s}}] \right)\!.
\end{equation}
For identical molecules, the plasmon thus induces an effective coupling of the
center-of-mass mode to a heat bath with coupling rate $\gamma_c = 4|\alpha|^2
g_1^2 \Re(S_{-1} - S_{1})$ and occupation number $\bar n_c = \Re{S_1} /
\Re(S_{-1} - S_{1})$, corresponding to an effective bath temperature of
$k_\mathrm{B} T_c = \omega_1/ \log\left(\Re S_{-1}/\Re S_1\right)$. 

\section{Results}\label{sec:results}
\subsection{Identical molecules}\label{sec:sym}

\begin{figure}
  \includegraphics[width=\linewidth]{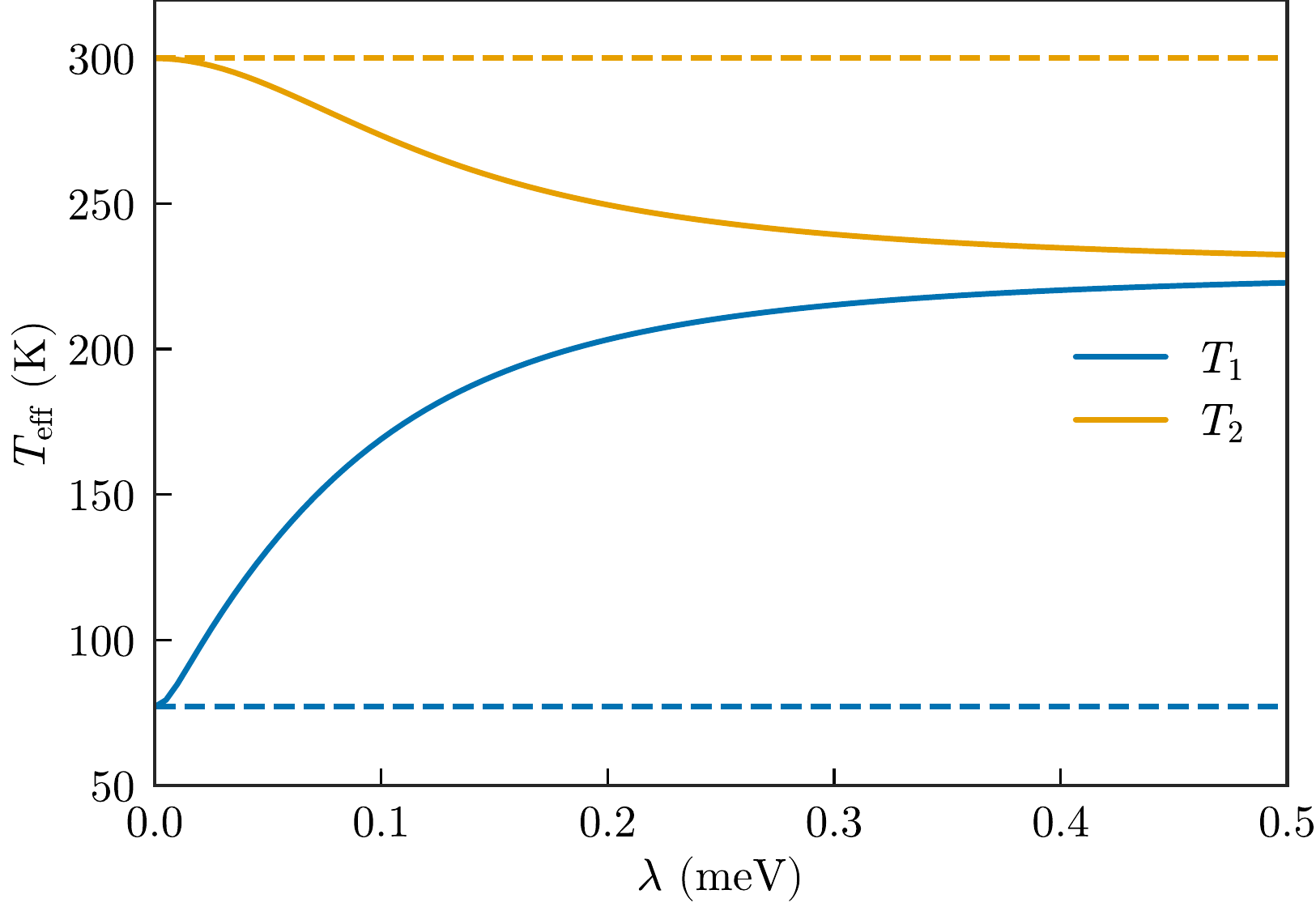}
  \caption{Effective temperature $T_\text{eff}$ of the molecules as a function
  of the dipole-dipole coupling constant $\lambda$ when coupling to the plasmon
  mode is negligible ($g_{1} = g_{2} = 0$). Dashed lines indicate the
  temperature of the bath for each molecule.}
  \label{fig:2}
\end{figure}

For reference, in \autoref{fig:2} we first show the heat transfer due to the
direct dipole-dipole interaction $\lambda$ between the molecules outside of a
cavity, i.e., when there is no interaction with the plasmon mode. Here and in
the following, we choose phonon mode frequencies of $\omega_{1} = \omega_{2} =
50$~meV, with the external baths at temperatures of $T_{1} = 77$~K and $T_{2} =
300$~K, and molecule-bath coupling rates given by $\gamma_{1} = \gamma_{2} =
0.25$~meV. Not surprisingly, as $\lambda$ is increased, the molecules exchange
energy more efficiently, causing heat to flow between them and their effective
temperatures to approach each other. When $\lambda$ becomes comparable to
$\gamma_1 = \gamma_2$, i.e., energy exchange between the molecules is comparably
fast to the molecule-bath coupling, the steady-state temperatures of the two
molecules become almost equal. We also note that due to the symmetry of the
system in this case, the change in mean phonon numbers (not shown) induced by
the coupling is symmetric, $\delta n_{1} = - \delta n_{2}$, such that the total
phonon number in both molecules is conserved as $\lambda$ is increased.

\begin{figure}
  \includegraphics[width=\linewidth]{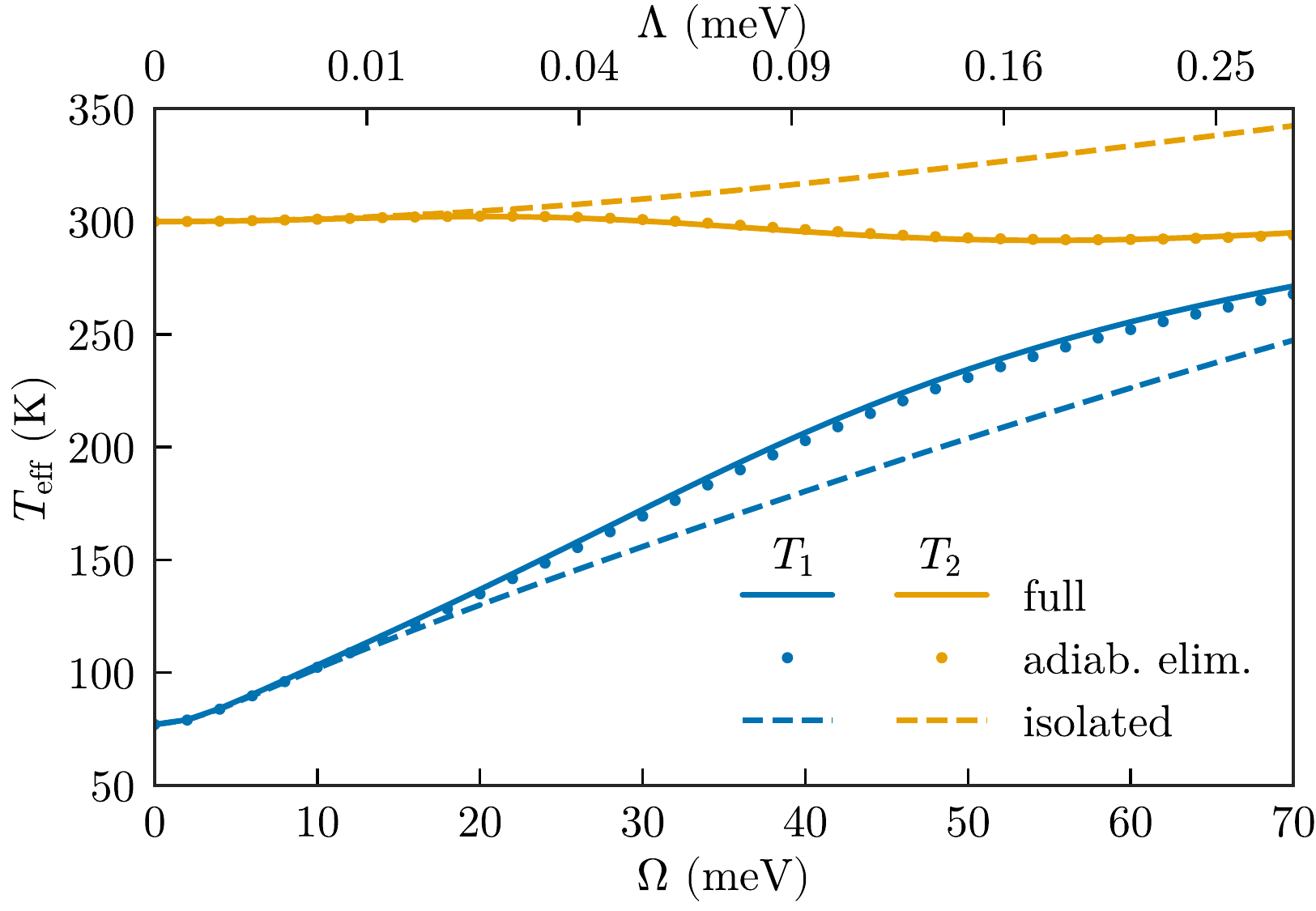}
  \caption{Effective temperature of two driven coupled molecules as a function
  of driving strength $\Omega$, compared to the case of isolated molecules.
  Solid lines show results of the full numerical simulations, while dots show
  the results obtained after adiabatic elimination and applying the RWA\@.
  Dashed lines show the equivalent results for the case of isolated molecules.
  In all cases, blue (light orange) lines and symbols correspond to the colder
  (hotter) molecule 1 (2). The upper axis shows the values of the effective
  plasmon-induced molecule-molecule coupling strength $\Lambda$ obtained through
  adiabatic elimination.}
  \label{fig:3}
\end{figure}

\begin{figure*}
  \includegraphics[width=\linewidth]{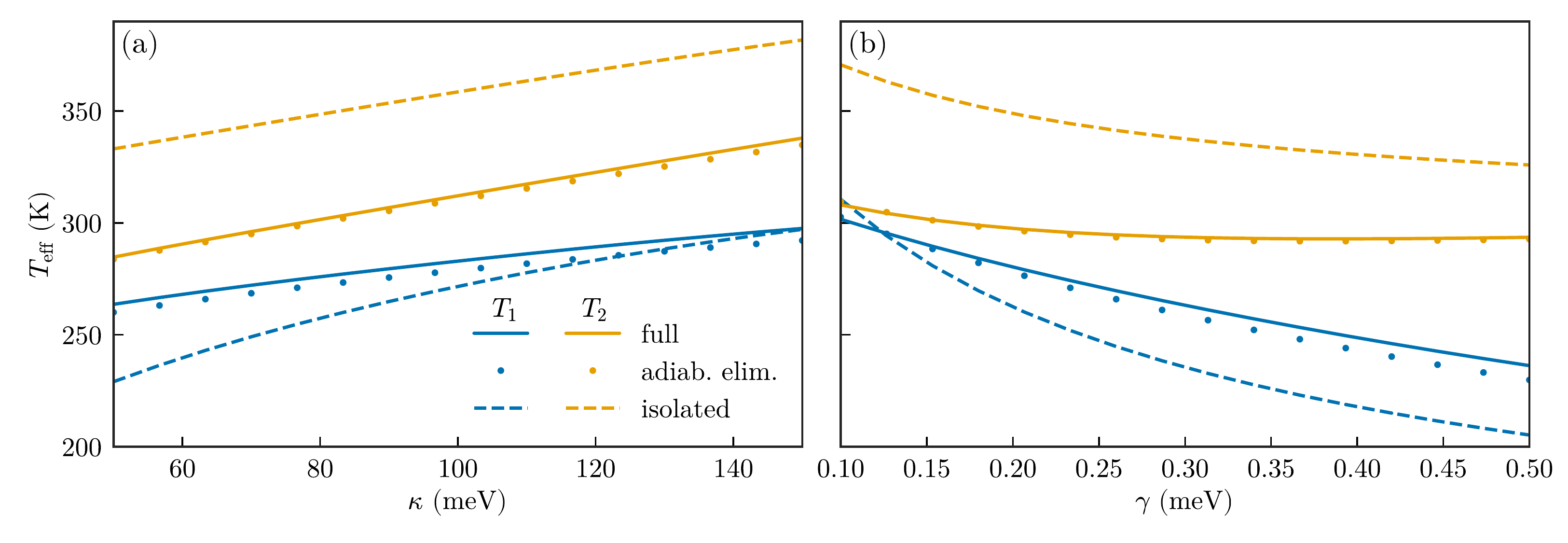}
  \caption{Effective temperature of the molecules as a function of (a) the
  cavity damping rate $\kappa$ and (b) the molecular damping rate
  $\gamma=\gamma_1=\gamma_2$. In both panels, all other parameters are kept
  constant at the values given in the main text. The color and line styles are
  identical to \autoref{fig:3}.}
  \label{fig:4}
\end{figure*}

By comparison, in \autoref{fig:3} we study the case where there is no direct
dipole-dipole interaction, but the optomechanical coupling to the plasmon is
nonzero, and the plasmon mode is driven by an external driving laser. We use a
plasmonic resonance frequency of $\omega_{c} = 1.36$~eV with linewidth $\kappa =
68$~meV, corresponding to a quality factor of $Q=20$. The optomechanical
coupling rate is taken as $g_{1} = g_{2} = 10$~meV, similar to values derived in
the literature~\cite{Roelli2016, Schmidt2016Quantum, Schmidt2017}, and the
laser-plasmon detuning is set to $\delta_{0} = 150$~meV. Due to the dispersive
nature of the plasmon-phonon interaction, the plasmon mode does not have any
influence on the phonon population when there is no driving, since in that case
$\langle \hat{a}^\dagger \hat{a}\rangle = 0$. When the external laser is turned
on, the molecules can be driven out of equilibrium with their local heat baths.
This leads to two possible effects on the molecular temperature: On the one
hand, vibrational pumping of phonons through Stokes (anti-Stokes) transitions
can heat (cool) the molecules~\cite{Schmidt2017}. This is a well-known
single-molecule effect that also occurs when each molecule is alone in the
cavity, as shown in dashed lines in \autoref{fig:3}. On the other hand, the
effective molecule-molecule interaction mediated by the plasmon additionally
enables heat transfer between the molecules, leading the molecular temperatures
(solid lines in \autoref{fig:3}) to approach each other compared to the
single-molecule case, even though there is no direct molecule-molecule
interaction ($\lambda=0$). Noticeably, this coupling becomes large enough to
even reverse the trend in the change of the temperature of the hotter molecule:
Although it gets heated by the plasmon when it is in the cavity by itself, its
temperature \emph{decreases} in the presence of the colder molecule due to their
effective coupling induced by the plasmon. The results obtained using
\autoref{eq:master_adiab_RWA}, i.e., adiabatic elimination within the RWA, are
shown as dots in \autoref{fig:3}, and are found to agree well with the full
numerical solution.

The good agreement between the full calculation and the adiabatic elimination
procedure permit a more in-depth analysis of the results by examining the
obtained effective coupling parameters. We therefore also show the corresponding
value of $\Lambda$, obtained from \autoref{eq:Lambda}, as the upper $x$-axis of
\autoref{fig:3}, demonstrating that the observed values become large enough to
induce significant energy transfer between the vibrations in the two molecules.
We also note that for the parameters used here, the effective temperature of the
common bath is almost independent of $\Omega$, with a value of $T_c\approx
444~$K (such that the molecules are heated by the plasmon-enhanced laser
driving), while both the bath coupling rate $\gamma_c$ and the effective
molecule-molecule interaction $\Lambda$ are to a good approximation proportional
to $\Omega^2$, with $\gamma_c \approx 0.64 |\Lambda|$.

\begin{figure*}
  \includegraphics[width=\linewidth]{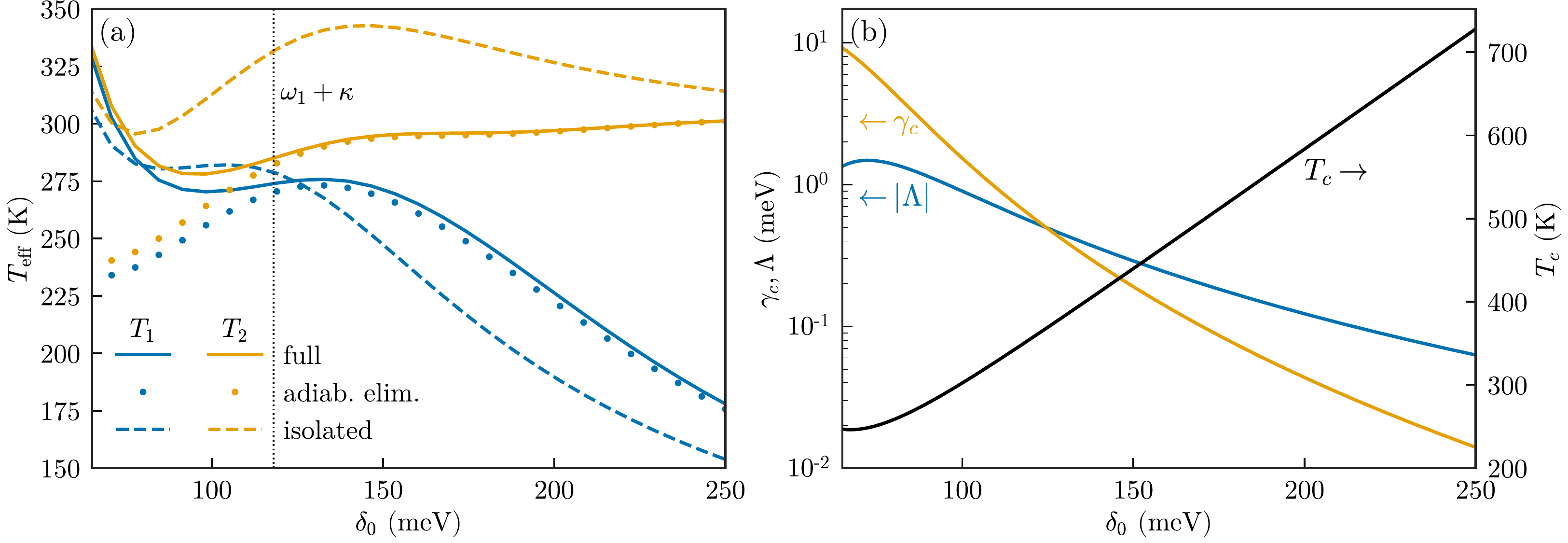}
  \caption{(a) Effective temperature of the molecules as a function of the
   laser-plasmon detuning $\delta_0$ (same colors and line styles as in
   \autoref{fig:3}). (b) Parameters obtained from adiabatic elimination of the
   plasmon mode: molecule-molecule coupling $\Lambda$, coupling rate $\gamma_c$
   of the center-of-mass mode to its bath, and new bath temperature $T_c$. In
   both panels, all other parameters are kept constant at the values given in
   the main text.}
  \label{fig:5}
\end{figure*}

We next study the influence of various parameters on the results obtained above,
setting the driving strength to $\Omega = 70~$meV (the largest value considered
in \autoref{fig:3}) here and later. The effectiveness of the plasmon-mediated
heat transport, as measured by the deviation of the full results from those with
each molecule by itself in a cavity, is reduced when the cavity loss rate
$\kappa$ increases (keeping all other parameters constant), as shown in
\autoref{fig:4}(a). Similarly, increasing the molecule-bath coupling
$\gamma_1=\gamma_2$ leads to more efficient thermalization of each molecule with
its individual bath, such that their temperatures approach those of their baths,
as seen in \autoref{fig:4}(b). However, plasmon-mediated heat transfer still
constitutes an important channel and leads to significant deviations between the
individual-molecule results and the coupled system. We note that changing the
optomechanical coupling rates $g_{1}=g_{2}$ (not shown) is equivalent to
changing the external driving $\Omega$ (see \autoref{fig:3}), and leads to more
efficient energy transfer (and also more efficient heating).

Finally, we investigate the effect of the laser-plasmon detuning $\delta_0$.
Since the coherent and incoherent interactions induced by the plasmon within
adiabatic elimination depend on the imaginary and real part of the function
$S(\omega)$, respectively, it can be be anticipated that their behavior as a
function of $\delta_0$ is quite different. As shown in \autoref{fig:5}(a), this
leads to a relatively complex dependence of the results on the laser detuning,
even for the case of $\delta_0 > \omega_1$ that we are focusing on here. First
of all, it can be seen that the adiabatic elimination procedure only works well
for large enough detunings, essentially when $\delta_0 > \omega_1 + \kappa$
(indicated by a thin dotted line in \autoref{fig:5}a). For smaller values, the
vibrational mode is quasi-resonantly pumped by the laser, with the effective
temperature increasing strongly. For larger values of the detuning, the
temperature $T_1$ of the colder molecule decreases noticeably, while $T_2$ stays
approximately constant and only increases slowly. This behavior can be
understood by studying the effective parameters, shown in \autoref{fig:5}(b).
This shows that both the effective molecule-molecule coupling $\Lambda$ as well
as the coupling rate $\gamma_c$ to the common bath decrease as $\delta_0$
becomes larger, with $\Lambda$ having a longer tail. One could thus conclude
that larger detunings could be used to increase the relative importance of the
coherent molecule-molecule interaction (and thus direct energy transfer between
the molecules) compared to the coupling $\gamma_c$ to the common bath, while
increasing the driving intensity $\Omega$ to maintain the same absolute value of
$\Lambda$ (both $\gamma_c$ and $\Lambda$ scale with $\propto\Omega^2$). However,
this strategy is rendered ineffective by the concomitant increase in the common
bath temperature $T_c$ for larger $\delta_0$, as seen in \autoref{fig:5}(b),
such that the overall heating of the molecules stays similarly efficient for
different values of $\delta_0$.

\subsection{Non-symmetric system}\label{sec:asym}

\begin{figure}
  \includegraphics[width=\linewidth]{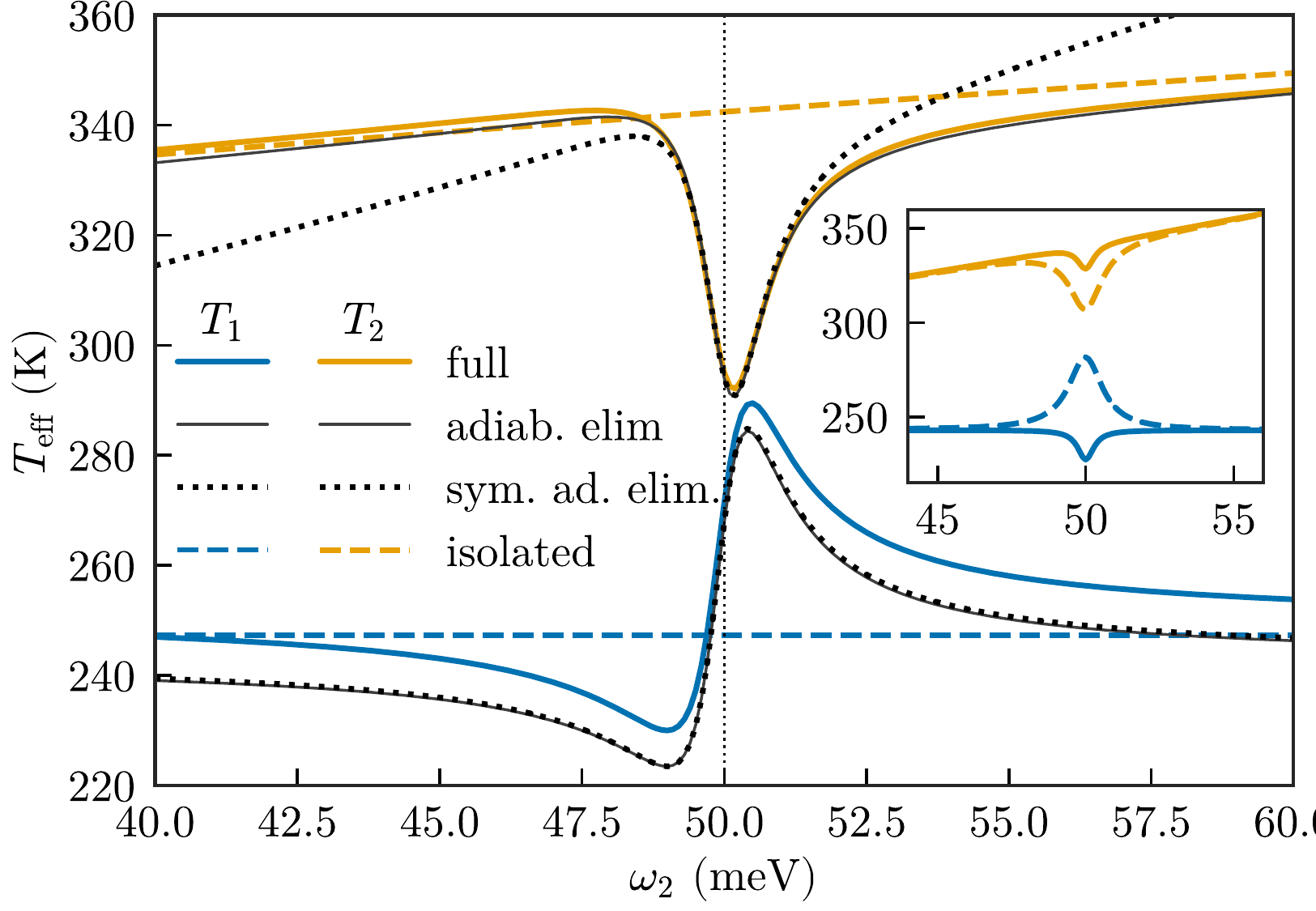}
  \caption{Effective temperature of both molecules as a function of the
  frequency $\omega_2$ of the hotter molecule. The results obtained after
  adiabatic elimination of the cavity mode are shown both with the actual
  frequencies (thin black lines), as well as under the symmetric approximation
  $\omega_1=\omega_2$ (dotted black lines). The inset shows the results obtained
  within adiabatic elimination when either the coherent interaction $\Lambda$ or
  the incoherent interaction (terms mixing $b_1$ and $b_2$ in
  \autoref{eq:Lad_RWAsym}) is removed (solid and dashed lines, respectively).}
  \label{fig:6}
\end{figure}

In this section, we investigate the non-symmetric situation where two molecules
with different vibrational mode frequencies are coupled to the same plasmonic
resonance. For simplicity, we only change the mode frequency of molecule $2$ and
keep all other parameters (couplings and bandwidths) constant, and thus
identical for both molecules. In \autoref{fig:6}, we show the temperature of
both molecules when changing the frequency of the hotter molecule, with all
other parameters as in \autoref{fig:3} (in particular, $\omega_1=50$~meV). It
can immediately be appreciated that plasmon-induced heat transfer between the
molecules is only efficient when the two vibrational modes are close to
resonance, with a central peak visible where the hotter (colder) molecule is
cooled (heated) compared to the single-molecule case. We note that the width of
these peaks is approximately determined by the overall broadening induced by
coupling to the different heat baths, both the individual baths of each molecule
as well as the effective common heat bath created by the plasmon (see
\autoref{sec:adiabatic_elimination}).

Interestingly, while the hotter molecule displays an almost Lorentzian-like
lineshape, i.e., it is more efficiently cooled the closer the two molecules are
to resonance, the colder molecule shows a Fano-like lineshape as a function of
frequency difference. Its temperature actually decreases below the
single-molecule value at the same driving when the hotter molecule is at a
slightly lower frequency than the colder one. As seen in \autoref{fig:6}, this
behavior is well-reproduced using adiabatic elimination (thick gray lines), and
is also captured when doing the additional ``symmetric approximation'' that
$\omega_1=\omega_2$ within the terms induced by adiabatic elimination, shown as
dotted black lines. This asymmetric lineshape can then be shown to occur due to
the competition of two separate effects: On the one hand, the direct (coherent)
molecule-molecule coupling mediated by the plasmon leads to more efficient
energy transfer when the two molecules are close to resonance, and also induces
an energy shift on the symmetric mode. On the other hand, the incoherent
coupling of the two molecules to a common heat bath becomes less efficiently
when they are on resonance. This is demonstrated in the inset of
\autoref{fig:6}, where the solid lines show the results of adiabatic elimination
if the coherent coupling $\Lambda$ is set to zero, while the dashed lines show
the results when mixed terms containing $b_1$ and $b_2$ are removed from the
incoherent contribution $L_{\mathrm{ad}}^{\mathrm{sym}}[\rho_{\mathrm{s}}]$ in
\autoref{eq:Lad_RWAsym}. The combination of these two effects leads to an
asymmetric line shape with regions where the plasmon-induced contribution heats
the colder molecule less efficiently than in the isolated case, even though it
is additionally effectively coupled to the hotter molecule.

\begin{figure*}
  \includegraphics[width=\linewidth]{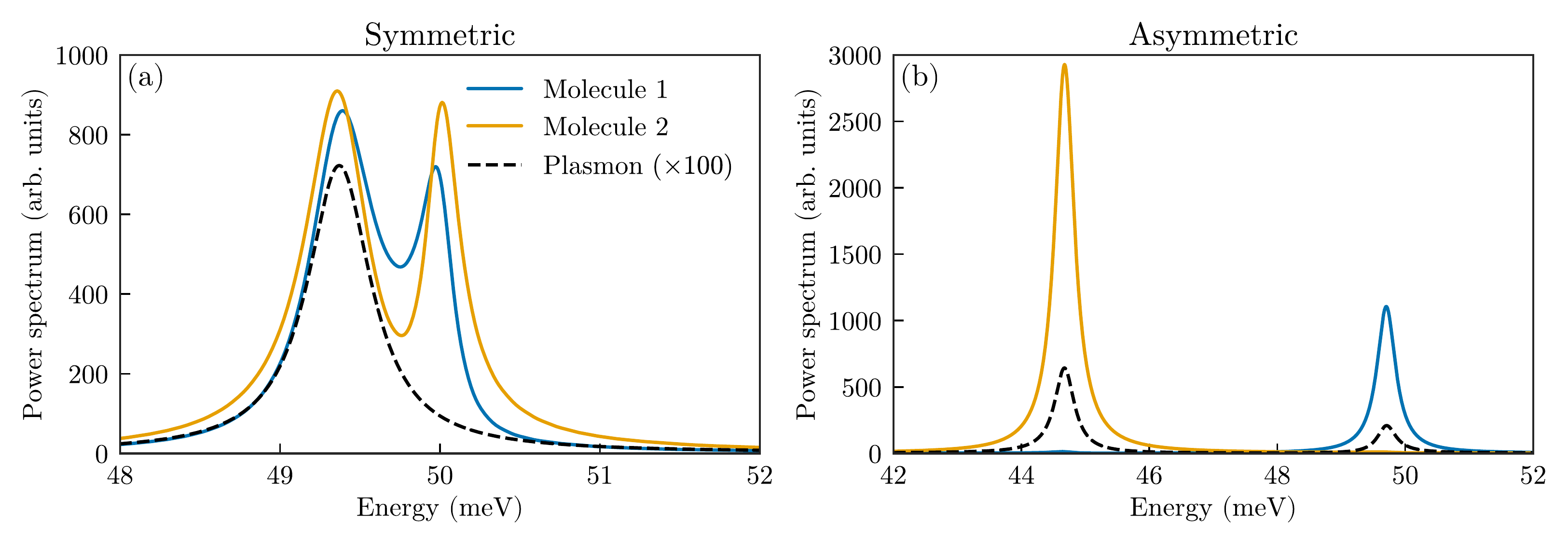}
  \caption{(a) Power spectrum density for the cold molecule (blue), hot molecule
  (light orange) and plasmon mode in symmetric system (dashed black). (b) Power
  spectrum density for non-symmetric system.}
  \label{fig:7}
\end{figure*}
\begin{figure*}
  \includegraphics[width=\linewidth]{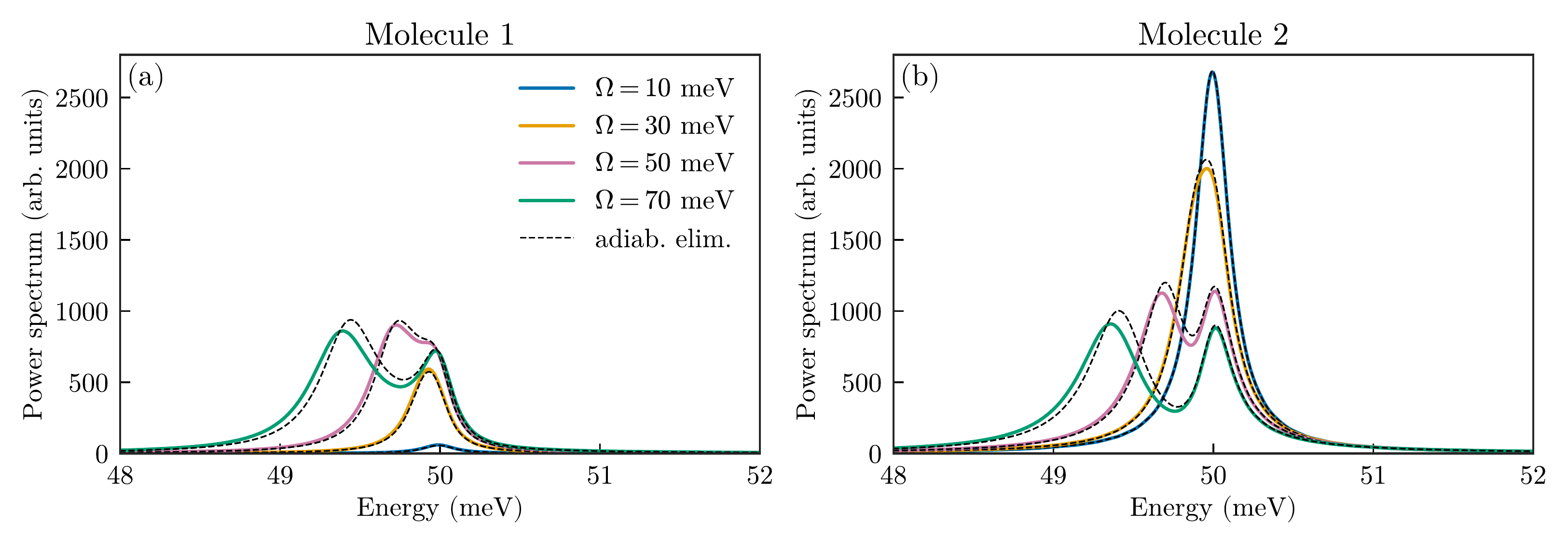}
  \caption{Power spectrum density in the symmetric system ($\omega_1 = \omega_2
  = 50~$meV), for different driving strengths $\Omega$. As the driving strength
  is increased, the vibrational modes of the two molecules enter into strong
  coupling with a double-peaked structure, i.e., an effective Rabi splitting,
  visible both for the (a) cold molecule and (b) hot molecule. Dashed black
  lines show the results obtained after adiabatic elimination of the cavity mode
  and applying the RWA.}
  \label{fig:8}
\end{figure*}

\subsection{Power spectral density}\label{sec:psd}
We next study the power spectral density of the molecular vibrations and the
plasmon, which gives additional insight into the dynamics of the system by
showing the effective oscillation frequencies of the various components. The
PSD, defined as
\begin{equation}\label{eq:PSD}
  S_{\hat{c}}(\omega) = \int_0^\infty e^{-i\omega t} \langle \hat{c}^{\dagger}(t)\hat{c}(0) \rangle_{\text{ss}} \mathrm{d}t,
\end{equation}
gives a measure of the oscillation spectrum of a mode as a function of
frequency~\cite{Gardiner2004}. For the symmetric case of identical molecules,
\autoref{fig:7}(a), both molecules show a double-peaked spectrum, corresponding
to mode splitting between the center-of-mass mode and the difference mode of the
molecular vibrations induced by the plasmon. This demonstrates clearly that the
effective coupling between the vibrational modes induced by the plasmonic mode
becomes large enough for the parameters used here that normal mode splitting
between the two vibrational modes occurs. Furthermore, it can be clearly
observed that the plasmonic resonance is only modulated at the frequency of the
center-of-mass mode, as expected from the discussion in
\autoref{sec:adiabatic_elimination}, with a clear red-shift compared to the bare
molecular vibrations.

In contrast, when the two molecular vibrations have sufficiently different
frequencies, as shown in \autoref{fig:7}(b) for $\omega_1=50$~meV,
$\omega_2=45$~meV, they each induce a separate modulation onto the plasmonic
mode, and each molecule only is influenced by the plasmonic mode modulation at
its own frequency, such that no effective coupling between molecules takes
place. In this figure, it can also be appreciated that the hot molecule induces
much stronger fluctuations on the plasmonic mode than the cold one.

In order to understand the onset of normal mode splitting between the molecules,
we plot the PSD of the two molecules for various values of the driving intensity
$\Omega$ in \autoref{fig:8}. As can be seen, for weak driving, the two molecules
only oscillate at their natural frequency, with much weaker excitation of the
cold molecule compared to the hot one. However, as $\Omega$ is increased above
about $30~$meV, the driving of the plasmonic mode induces a large enough
effective coupling between the molecules to overcome losses and lead to normal
mode splitting (or ``strong coupling'') between the vibrational modes. The
accordingly efficient energy transfer between the molecules then also leads to
much more similar amplitudes for the two molecular oscillations. The results
obtained within the adiabatic elimination are included in \autoref{fig:8} as
dashed black lines, again showing good agreement with the full numerical
results. Interestingly, this also shows that the quality of this approximation
actually decreases with increasing driving, and the splitting observed in the
full results is slightly larger than predicted by adiabatic elimination.

\section{Summary \& Discussion}\label{sec:summary}
To summarize, we have demonstrated that mutual coupling of two molecular
vibrations to a localized surface plasmon resonance in the optomechanical regime
can lead to efficient plasmon-mediated heat transfer between the molecules.
Importantly, this plasmon-mediated channel is only active when the plasmonic
mode is driven by an external laser field. This could enable active control of
heat transfer between molecules through an external laser field. Additionally,
in some parameter regimes, this optomechanically induced plasmon-mediated heat
transfer is more efficient than bare plasmon-induced heating, such that the
hotter molecule can effectively be cooled even though it is actively heated by a
relatively intense laser pulse. This is reminiscent of radiative cooling under
sunlight~\cite{Raman2014}, with the additional twist that here it is the
external laser field itself that induces the cooling by opening a new heat
transport channel to a colder reservoir.

Furthermore, we have shown that in the case of non-identical molecules, heat
transfer only efficiently takes place if the relevant vibrational modes are
close to resonance with each other, which can be understood as each molecule
only interacting with modulations of its own frequency imprinted on the plasmon
mode. We note that since \autoref{eq:hamiltonian} also could be used to describe
two vibrational modes of a single molecule, our results also imply that
different vibrational modes should behave essentially independently as long as
their frequencies are well-separated compared to their linewidths, and thus
provides some additional justification for the common use of single-mode
models~\cite{Roelli2016,Benz2016,Schmidt2017,Lombardi2018}.

Deeper insight is gained through the analytic approach of adiabatic elimination,
which reveals both coherent and incoherent coupling terms induced between the
molecular vibrations by the plasmon mode. For example, this reveals that
inducing a slight detuning between the vibrational frequencies can lead to a
competition between the coherent and the incoherent plasmon-induced coupling
terms that leads to a Fano-like lineshape where the coupling to the hotter
molecule still cools down the colder molecule compared to the case where it is
coupled to the laser-driven plasmon mode by itself.

\begin{acknowledgments}
This work has been funded by the European Research Council (ERC-2016-STG-714870)
and the Spanish MINECO under contract MAT2014-53432-C5-5-R and the ``María de
Maeztu'' programme for Units of Excellence in R\&D (MDM-2014-0377), as well as
through a Ramón y Cajal grant (JF) and support from the Iranian Ministry of
Science, Research and Technology (SMA).
\end{acknowledgments}

\bibliography{references}

\end{document}